\documentclass[11pt,twoside]{article}

\usepackage{asp2006}
\usepackage{lscape}

\usepackage{graphicx}
\usepackage{sidecap}
\usepackage{adassconf}
\usepackage{threeparttable}

\begin{document}
\title{Spitzer's View of Edge-on Spirals}   
\author{B. W. Holwerda\altaffilmark{1}, R. S. de Jong\altaffilmark{1}, A. Seth \altaffilmark {2}, J. J. Dalcanton \altaffilmark{3}, M. Regan \altaffilmark{1}, E. Bell \altaffilmark{4} and S. Bianchi \altaffilmark{5}}

\altaffiltext{1}{Space Telescope Science Institute} 
\altaffiltext{2}{Center for Astrophysics}
\altaffiltext{3}{University of Washington}
\altaffiltext{4}{Max Planck Inst. Heidelberg}
\altaffiltext{5}{Istituto di Radioastronomia / CNR}


\begin{abstract} 
Edge-on spiral galaxies offer a unique perspective on disks. One can
accurately determine the height distribution of stars and ISM and the
line-of-sight integration allows for the study of faint structures. The
Spitzer IRAC camera is an ideal instrument to study both the ISM and
stellar structure in nearby galaxies; two of its channels trace the old
stellar disk with little extinction and the 8 micron channel is dominated
by the smallest dust grains (Polycyclic Aromatic Hydrocarbons, PAHs).
\cite{Dalcanton04} probed the link between the appearance of dust
lanes and the disk stability. In a sample of bulge-less disks they show
how in massive disks the ISM collapses into the characteristic thin dust
lane. Less massive disks are gravitationally stable and their dust
morphology is fractured. The transition occurs at 120 km/s for bulgeless
disks.
Here we report on our results of our Spitzer/IRAC survey of nearby edge-on
spirals and its first results on the NIR Tully-Fischer relation, and ISM disk stability.
\end{abstract}





\section{Introduction}

For 32 edge-on spiral galaxies, spanning Hubble type and mass, we fit the edge-on infinite disk model by \cite{vdKruit81a} on the IRAC mocaics: $\mu(R, z) = \mu (0,0) ~ \left({R/h}\right)K_1\left({R/h}\right) ~ sech^2\left({z/z_0}\right)$. We applied stellar dominated 3.6 and 4.5 $\mu$m  channels and the PAH emission at 8.0 $\mu$m, with the stellar contribution subtracted \citep[using the prescription from ][]{Pahre04a}. The data is from our dedicated GO program (GO 20268) and the {\em Spitzer} archive. Our program aims to populate the lower range of disk masses. 
From the fit we obtain the scale-length ($h$), the scale-height ($z_0$), and the face-on central surface brightness ($\mu_0$). The edge-on disk model fits the stellar channels very well but the PAH channel less so because of star-formation structures, such as HII regions, in the PAH maps.
The disk's total luminosity is inferred from the fitted model: $L_{disk} = 2 \pi h^2 \mu_0$, with $h$ the scale-length and $\mu_0$ the face-on central surface brightness \citep[See][]{Kregel02}.
Combined with the rotational velocities from HyperLeda, we can construct the Tully-Fisher relation for our galaxy disks and a plot of disk oblateness as a function of rotational velocity (and hence dynamical mass).
%
%

\section{Tully-Fisher at 4.5 $\mu$m.}

\begin{figure}
  \centering
  \includegraphics[width=0.8\textwidth]{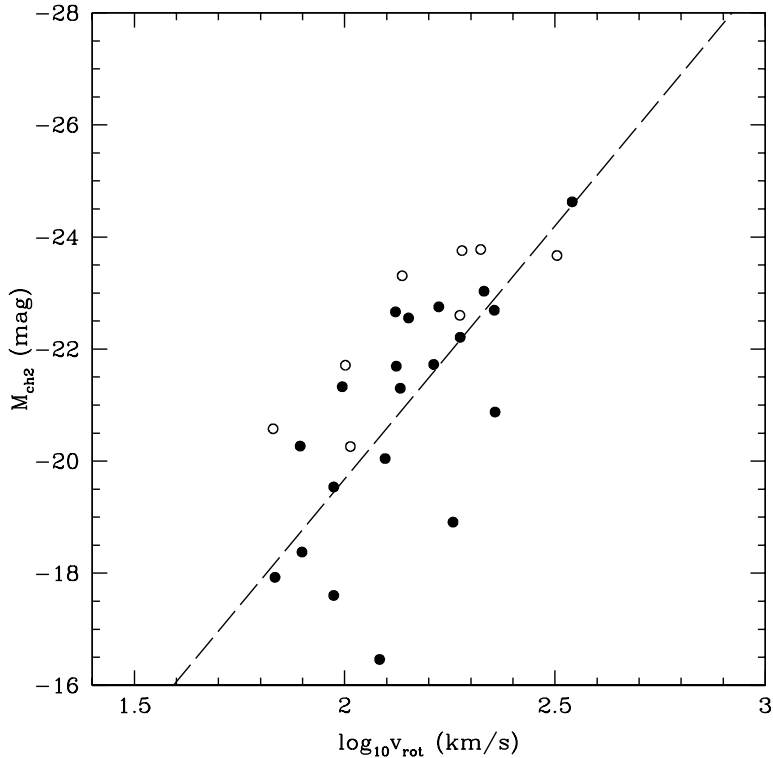}
\caption{\label{f:tf} The Tully-Fisher relation for disks in our sample. Notably, the slope is very similar to the one found by \cite{Meyer06a} for the SINGS sample. Open circles are the poorer disk fits ($\chi^2 > 10$).}
\end{figure}

Figure \ref{f:tf} shows the inferred Tully-Fischer relation for these disks in one of the stellar channels, 4.5 $\mu$m. Notably, the slope ($\alpha$) is 3.5, similar to what \cite{Meyer06a} found but contrary for the increasing trend of slope with redder filters. The effects of age and metallicity of the stellar population become independent and opposite effects on the color-M/L relation in NIR \citep[See][Figure 2d]{BelldeJong}. 
Hence, the shallower slope in the  IRAC stellar channels, could well be this metallicity effect starting to dominate.


\section{Oblateness of Stellar and PAH disks}

\begin{figure}
  \centering
 \includegraphics[width=0.8\textwidth]{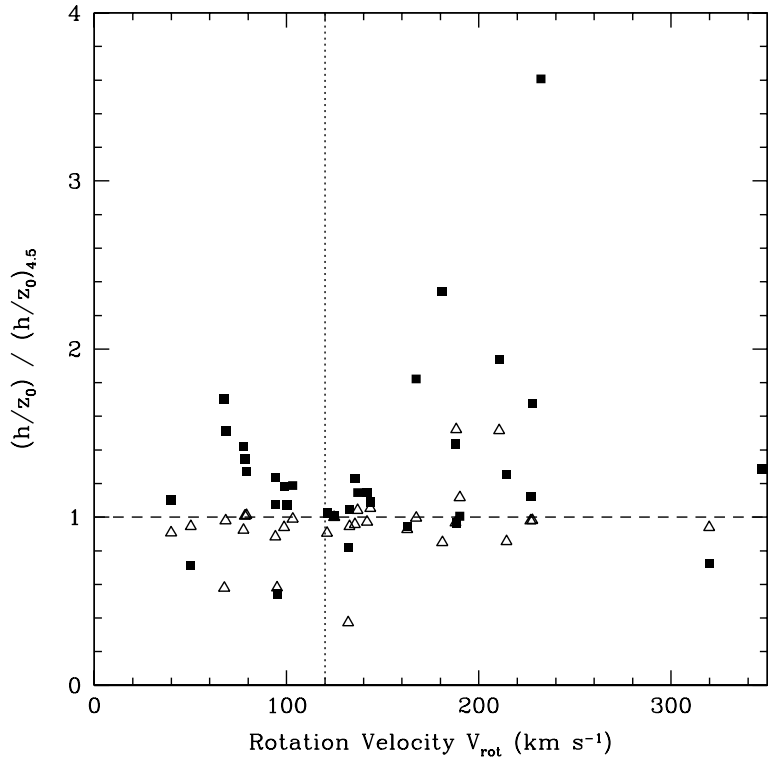}
\caption{\label{f:hz} The oblateness ($h/z_0$) of the 3.6 $\mu$m fit (open triangles) and 8.0 $\mu$m PAH map (filled squares) in terms of the 4.5 $\mu$m. A similar oblateness as the 4.5 $\mu$m. The oblateness of the stellar filters, 3.6 and 4.5 $\mu$m is very similar. The PAH disk is flatter than the stellar, especially in massive disks ($v_{rot} > 120 ~ km/s$, vertical dotted line).}
\end{figure}

Figure \ref{f:hz} shows the relation between the disk oblateness --the ratio of scale-length and -height, $h/z_0$-- relative to the oblateness in 4.5 $\mu$m. The 3.6 and 4.5 $\mu$m channels both trace the stellar population and their oblateness measurements agree well. The PAH mosaic (8 $\mu$m), shows flattening compared to stellar disk: $h/z_0 (PAH) < h/z_0 (4.5 \mu m.)$. The PAH disk's flattening is substantial in the case of the more massive disks ($v_{rot} > 120 ~ km/s$). This appears to corroborate the reasoning of \cite{Dalcanton04} that massive disks are vertically unstable and hence form the characteristic dust lanes.

\section{Conclusions}

Hence, from our disk models of IRAC data, we conclude: 

\begin{itemize}
\item[1] The Tully-Fisher relation has a shallower slope ($\alpha=3.5$) than naively expected from trends with filter (Figure \ref{f:tf}), which may be a metallicity effect of the stellar population's M/L.
\item[2] The PAH disk is flatter than the stellar one, especially in more massive spirals (Figure \ref{f:hz}). The effect is not as pronounced as found by \cite{Dalcanton04} and \cite{Obric06} based on dust lanes.
\end{itemize}

\section{Future Work}

The unique perspective of {\it Spitzer} on edge-on spirals allows us to quantify the vertical structure in spiral disks. 
The PAH maps are the product of both the ISM density and star-formation. The stellar channels are effectively unaffected by extinction and can therefore serve as a reference to quantify dust lanes and smaller dust extinction structures. The phenomenon of truncation --the sudden change in exponential behavior of disks-- can also be characterized for all IRAC bands.



\end{document}